\documentclass[a4paper]{spie} 
\usepackage[]{graphicx}
\usepackage{ifthen}

\usepackage{xspace}
\usepackage{array}
\usepackage{changes}
\usepackage{amsmath,amsfonts,amssymb}
\usepackage[colorlinks=true, allcolors=blue]{hyperref}
\usepackage{color}
\newcommand{\pathpic}{} 

\newcommand{\jpgext}{jpg}

\newcommand{\figtable}[4]  {
  	 \begin{figure}[ht]
          \begin{minipage}[t]{0.48\linewidth}\centering
          \includegraphics[width=\textwidth]{pic/\pathpic/#1.\jpgext}
          \caption{#2 }\label{fig.\pathpic#1}
          \end{minipage}
          \hspace{0.2cm}
          \begin{minipage}[t]{0.48\linewidth}   \centering
          \includegraphics[width=\textwidth]{pic/\pathpic/#3.\jpgext}
          \caption{#4}\label{fig.\pathpic#3}
          \end{minipage}

      \end{figure}
  }

\newcommand{\fig}[3]  {

        \vspace{0.2 cm}
        \begin{figure}[h]
        \begin{center}
        \includegraphics[height=#2cm]{pic/\pathpic#1.\jpgext}
        \end{center}\vspace{-0.4cm} \caption{#3}
        \label{fig.\pathpic#1}
        \end{figure}
                       }

\newcommand{\verify}[1]{   \ifthenelse{\boolean{accept}}{#1}{\textcolor{red}{#1 (to be verified!)}}               }

\newcommand{\um}{$\mu$m\xspace}

\title{Contactless actuators and pyramid wavefront sensor, the SPLATT concept for space active optics: an overview of the project and the last laboratory results.  }
\author[a,i]{Runa Briguglio}
\author[a,i]{Marco Xompero}
\author[b]{Marcello Scalera}
\author[b]{Marco Riva}
\author[a,i]{Ciro Del Vecchio}
\author[a,i]{Luca Carbonaro}
\author[c,i]{Carmelo Arcidiacono}
\author[a,i]{Guido Agapito}
\author[a,i]{Enrico Pinna}
\author[d,i]{Alessandro Terreri}
\author[d,i]{Fernando Pedichini}
\author[e,i]{Riccardo Muradore}
\author[f]{Matteo Tintori}
\author[f]{Daniele Gallieni}
\author[g]{Roberto Biasi}
\author[g]{Christian Patauner}
\author[h]{Alessandro Zuccaro Marchi}

\affil[a]{INAF Osservatorio Astrofisico Arcetri, L. E. Fermi 5, 50125 Firenze Italy }
\affil[b]{INAF Osservatorio Astronomico di Merate, Via E. Bianchi, 46, 23807 Lecco Italy}
\affil[c]{INAF Osservatorio Astronomico di Padova, Vicolo dell'Osservatorio, 5, 35122 Padova Italy}
\affil[d]{INAF Osservatorio Astronomico di Roma,  Via Frascati, 33, 00078 Monte Porzio Catone Roma Italy}
\affil[e]{Università di Verona, Via S. Francesco, 22, 37129 Verona Italy}
\affil[f]{A.D.S. - International via Pio Galli sindacalista, 3, 23841 Annone di Brianza, Lecco Italy}
\affil[g]{Microgate, Via Waltraud-Gebert-Deeg, 3e, 39100 Bolzano Italy}
\affil[h]{ESA ESTEC, Keplerlaan 1, PO Box 299,2200 AG, Noordwijk, The Netherlands}
\affil[i]{ADONI, Italian National Adaptive Optics Laboratory}

\authorinfo{Further author information:\\Runa Briguglio: E-mail: runa.briguglio@inaf.it, Telephone: +39 055 2752200\\}

\begin{document} 

  \maketitle

  \begin{abstract}
  In the last few years the concept of an active space telescope has been greatly developed, to  meet demanding requirements with a substantial reduction of tolerances, risks and costs. This is the frame of the LATT project (an ESA TRP) and its follow-up SPLATT (an INAF funded R\&D project). Within the SPLATT activities, we outline a novel approach and investigate,  both via simulations and in the optical laboratory, two main elements: an active  segmented primary with contactless actuators and a pyramid wavefront sensor (PWFS) to drive the correction chain. The key point is the synergy between them: the sensitivity of the PWFS and the intrinsic stability of a contactless-actuated mirror segment.
 Voice-coil, contactless actuators are in facts a natural decoupling layer between the payload and the optical surface and can suppress the high frequency vibration as we verified in the lab. We subjected a 40 cm diameter prototype with 19 actuators to an externally injected vibration spectrum; we then measured optically the reduction of vibrations when the optical surface is floating controlled by the actuators, thus validating the concept at the first stage of the design. 
  The PWFS, which is largely adopted on ground-based telescope, is a pupil-conjugated sensor and offers a user-selectable sampling and capture range, in order to match different use cases; it is also more sensitive than Shack-Hartmann sensor especially at the low-mid spatial scales. We run a set of numerical simulations with the PWFS measuring the misalignment and phase steps of a JWST-like primary mirrors: we investigated the PWFS sensitivity in the sub-nanometer regime in presence of photon and detector noise, and with guide star magnitudes in the range 8 to 14.
 In the paper we discuss the outcomes of the project and present a possible roadmap for further developments.
 
   \end{abstract}


\keywords{Space telescopes, optical stability, adaptive optics, LUVOIR, pyramid wavefront sensor, correction chain}
\section{Introduction}
    \subsection{Space telescopes and technological innovations}
    The ambitious science quests conceived during the past decade (e.g. cosmic origins, exploration of exoplanets) come together with demanding technological challenges. The space agencies developed therefore  a roadmap to \emph{fill the gap} between the current status and the future requirements. The LUVOIR studies (some of them in the references, plus the interim report) may be considered as a very good summary in terms of needs identification, overview of requirements, budgeting and expected impact on scientific performances. In particular, the requirements are extremely tight: the coronoagraphic imaging of exo-earths, for instance, has a preliminary error budget of 10 nm Wavefront error (WFE), 10 pm stability, $10^{-10}$ constrast. In such context, it is well acknowledged that the active control of the optical surfaces may result in a \emph{local} increased complexity, leading however to reduced \emph{global} cost and risks.\\
    Adaptive optics (AO) is a well established technology for ground based telescopes,  considered as a baseline (first light) for the Extremely 
    Large Telescopes. In an AO system a wavefront sensor (WFS) is illuminated by a guide star (natural or laser) to detect the aberrations induced by the atmosphere and command a deformable mirror (DM) to a shape minimizing the WFE. In the context of space telescopes the active correction would be used to correct the thermo-elastic deformations of the mirrors and restore the optical shape.\\
    Ground based AO deals commonly with complexity: large format DM, hundreds or thousands of actuators, fast loop frequency, high power and computational loads. The translation of such elements for a space application is not straightforward. As first, the driving requirements shall be identified. A crucial point here is to address them under the correct cost-benefit view: the goal is to reduce the global cost, risk and complexity. Here comes the ratio of this work: what do we have to offer, from the AO world, for an active space telescope?
    
    \subsection{LATT \& SPLATT: active primary mirrors for space}
    
    The LATT project (see \citeonline{icso2017}, \citeonline{latt2016-1}, and \citeonline{latt2016-2}) is an ESA-funded activity under a TRP grant, in the spirit of the technological developments depicted above. The idea is to leverage on the mature expertise in the field of adaptive optics (AO) for ground based telescopes and extend it to space. The goal in particular is to investigate the conversion of an adaptive secondary into an active primary, or a primary segment. The specific technology is that adopted for the deformable secondaries of the Large Binocular Telescope\cite{asm} (Arizona), the Magellan Telescope (Las Campanas, Chile, see \citeonline{magellan}) and Very Large Telescope (Cerro Paranal, Chile, see \citeonline{vlt-test}), and also for the Extremely Large Telescope adaptive M4 mirror\cite{m4} and the Giant Magellan Telescope adaptive M2 (both currently under construction). The concept is based on a thin Zerodur-glass shell shaped by voice-coil actuators, which are in turn controlled in local close-loop fed by co-located capacitive position sensors; the force produced by the actuators is then adjusted (at high frequency) to obtain the wanted displacement on the position sensor. When working in an adaptive optics system, such voice-coil DM are then controlled at two speeds: in the high frequency regime (kHz) the actuator force is controlled by the capacitive sensor to keep the optical surface in position; at the working frequency (the one selected for the AO loop) the mirror is controlled by the WFS commands. The adaptive secondaries have been in service since 2010 and offered great opportunities for science.\\
    In the LATT project, a demonstrator called OBB was fitted with 19 actuators and a 40 cm diameter spherical shell. The thin shell (TS) is 1 mm thick and made of Zerodur, with magnets bonded on its back and coupled with the coil mounted on a Reference Body (RB). The RB is made in aluminum honeycomb with a thin carbon fiber skin. The total areal density of the system is approximately $16 \, \, \mathrm{kg}/\mathrm{m}^2$, including the RB, the TS and the actuators electronic. Such value is particular attractive and could be further optimized with a design update. The system has been subjected to thermo-vacuum test and optical qualification. The concern was the optical controllability with such a low actuator counts, compared to active area; the test in the optical laboratory indicated that the final wavefront error (WFE) was consistent to that of much larger system. As a last point, the OBB was tested on the vibration stand where it was subjected to several vibration spectra, to simulate the launch conditions. The TS, which was the most concerning item, survived the test thanks to an electrostatic locking device, that has been designed specifically for the launch phase.\\
    
    \noindent The project ended in 2015 after the final review at ESA. During an informal follow up phase, we discussed the OBB features beyond the specific requirements and boundaries of the project. In particular we addressed:
    \begin{itemize}
        \item the correction ``geography'', i.e. primary vs post focal DM, or large vs small size DM;
        \item ways to further reduce the areal density through a design update;
        \item the implication of a large actuator count design (e.g. $\approx$ 1000 actuators on the primary), both from a power budget and and performance perspective;
        \item the architecture of the correction chain including the WFS.
    \end{itemize}
    A few points deserve some attention here. \\
    Voice coil motors (VCM) require a constant power supply, while set-and-forget actuators can be powered off after the correction. So the point is to compare \emph{globally} the benefits of VMC in the system. Are they a way to increase the system stability or performances?\\
    Thin shell technology (fabrication, handling, operation) is very mature, but would you really put an extremely fragile glass foil on a rocket? On the other side, what is the expected impact in terms of total mission weight and cost of an active primary with an areal density lower than $16 \, \, \mathrm{kg}/\mathrm{m}^2$, including WF control and support?
    The TS is the only mechanical component with optical requirements; the RB conversely can be manufactured to significantly lower specifications, since there is no mechanical contact between it and the TS. This permits also the use of extremely lightweight material such as aluminum honeycomb.\\
    
    \noindent Ground based AO made significant progresses in the field of WFS and control strategy: people there are particularly familiar with high orders (HO) systems, with thousands of actuators, and very large WFE due to air turbulence. Is there a WFS that best suits for space applications, also considering that it will operate at the diffraction limit? Does a dedicated WFS make a positive performance difference?\\
    Some of such questions have been addressed within an official follow up named SPLATT, Segments and Pyramids for Large Aperture Space Telescope Technology.

    \section{The SPLATT project}
    \subsection{Scenario and goals}
    The SPLATT project is an INAF activity funded under the INAF Tecno-PRIN2019,  and officially started in 2021. The team comes from the ADONI community, the italian network of people working in AO, including researchers both from scientific and technological field. Also from the specific technological point of view, the community is very heterogeneous due to a favourable mix of experts in optics, control theory, optical metrology, AIV, sensors and computer science. The SPLATT team gathered together people from 4 INAF departments, and shows a nice mix of competences. The idea is to build on such a mix to address the points raised above, in particular from the points of view of control, WFS and stability.\\
    The goal of the project is to increase the internal know how, by investigating the technological transfer from the ground based AO context, and by pursuing a system-level approach to the problems. In particular we considered two main aspects: the advantages offered by the contactless actuation provided by VCM and the high sensitivity of the pyramid WFS (PWFS, see \citeonline{CITA1}) operating in diffraction limited regime. Both points are to be discussed in the specific perspective of the LUVOIR studies, where optical stability is one of the key-points.\\
    The projects has three main tasks:
    \begin{itemize}
        \item install the OBB (which has been shipped back to INAF from ESA thanks to a loan agreement) on a dedicated optical bench, for further optical qualification;
        \item setup a segmented DM plus PWFS simulator, and explore the WF control stability in relevant conditions;
        \item draw a roadmap for further developments.
    \end{itemize}
    Let's analyze the first two items.\\
    The contactless actuation mechanism provided by VCM implies that there is no mechanical contact between the TS and the RB: the TS \emph{floats} at a given distance (or gap, typically 200 \um) in front of the RB. Several consequences can be derived: the manufacturing specs for the RB become less demanding; thermal deformations on the RB can be recovered with no fitting error by the actuators; mechanical vibrations from the spacecraft to the RB are not propagated to the optical surface, thus resulting in a significant gain in the optical stability. This is the point to be demonstrated in the laboratory: the vibration spectrum measured on the support is reduced  on the optical surface.\\
    
    \noindent The PWFS is a pupil-conjugated sensor, with a (double) glass prism installed at the telescope focus; the system provides a slope signal out of the 4 pupil images on the detector, where at least one pixel per sub-image is requested per mode-to-be-corrected. The PWFS has a non-linear, periodic response in the range $0<s<\lambda$ (where $s$ is the WF offset and $\lambda$ is the working wavelength). This point is a limitation on ground and is addressed by introducing a circular modulation; for space, this is conversely a valuable point to be exploited. In addition, the PWFS features an optical gain depending on the signal spatial scale. The highest sensitivity is for low spatial scale modes, which is a plus for segmented system since we expect to correct the segments mis-alignment.\\
    
    \noindent The first point, or the SPLATT experiment, will be discussed in Sec.\ref{sec.experiment} where we will present the laboratory setup. the test plan and the preliminary results. In Sec. \ref{sec.simulation} we will give an overview of the numerical code and present the simulation results.\\
    The project is currently (August 2022) at its mid-point; the activity in the laboratory  revealed that vacuum (or semi-vacuum) tests are mandatory for a further assessment of the OBB characterization within the specific scope of the project; we expect to deploy a low vacuum chamber by autumn 2022 and conclude the tests by the end of the year. The simulations are in turn basically completed: we plan therefore to start the project wrap-up, to prepare a development roadmap, by Spring 2023.
    
    \section{The SPLATT experiment}\label{sec.experiment}
        \subsection{The test plan and procedure}
        The test aims at demonstrating the vibration reduction when the TS floats in front of the RB thanks to the VCM. We need to compare two datasets to measure such a reduction: the first dataset is collected with the TS pulled against the RB by the VCM, so that to measure directly the oscillations of the mechanical structure; the second is with the TS floating. A dataset is composed by a large number (e.g. 2000) interferometric frames collected at high frequency (300 Hz). During the measurement, a disturbance is injected on the OBB to make the RB oscillate about the elevation axis; each frame is then analyzed to fit the tip/tilt, then the spectrum of the tilt time series is computed.\\
        In the end we have two time series and the associated spectra: the tip/tilt $T_p$ when the TS is pulled toward the RB, and the tip/tilt $T_f$ when the TS is floating. Since $T_p$ is a direct measurement of the disturbance injected, the ratio $T_f/T_p$ is the vibration attenuation.\\
        The disturbance is injected as a frequency sweep signal, in order to measure on a single dataset a broad frequency range (1 Hz to 120 Hz).
        \subsection{Laboratory setup}
        The test setup is composed by an optical bench with an interferometer  and the relay optics to illuminate the OBB; a flat mirror mounted at 45 $\deg$ steers the laser beam vertically toward the OBB. A separated test stand, sitting on the ground, holds the OBB above the optical bench. The disturbance is injected by means of a piezo actuators pushing the OBB elevation arm against a spring; the piezo is in turn fed by a waveform generator.  When the piezo is operated, the OBB stand oscillates about its elevation axis producing a vertical tilt as seen on the interferometer; the test stand is separated from the optical bench with the mirrors and interferometer to avoid propagating the vibrations on the other elements. The bench is also insulated from the ground to get rid of the environment noise. Such improvements came after a few weeks of commissioning of the test setup, in order to enhance the sensitivity.\\
        The interferometer is a PhaseCam6110 by 4D-Technology. It is a Twyman-Green dynamic interferometer and a phase map is produced out of a single frame captured, with typical exposure time of tens of microseconds. The CCD can work in cropped mode; we set a measuring area of 300x300 pixel on a 130 mm optical diameter on the OBB to get a frame rate of 320 Hz. Such high frame rate, together with a very short exposure time, allowed us to measure high frequency tilt signals on the OBB with no fringes smearing.\\
        The TS sits on a foam cushion for safety, placed at 5 mm distance from the RB. At the beginning of the testing session, the TS is lifted with the cushion to get in touch with the RB, then the actuators are powered to maximum negative (pulling) force to hold the TS against gravity; the cushion is then released to the safety distance.\\
        The OBB and test stand are fitted with accelerometers to measure the mechanical vibration injected by the piezo. One accelerometer, in particular, is placed at the outer edge of the OBB corresponding to the point of maximum oscillation. Its measurements may be integrated and compared with the interferometer signal.
        
        \figtable{20220810-134627}{OBB installed on its test stand.}{20220810-134747}{The TS sitting on the safety foam cushion. The yellow disk (coated with kapton) is the aluminum honeycomb RB. Please note the small air gap between the TS and the RB.}
        \figtable{20220810-134725}{Detail of the retention ring, of vertical support for the safety foam cushion.}{20220810-134718}{The piezo actuator installed on the elevation arm to inject the external disturbance.}

        \subsection{Preliminary results in air}
        As a very first step, we measured the SPLATT attenuation in a discrete frequency range 1 Hz to 120 Hz. The result is shown in Fig.~\ref{fig.response}, for different proportional gains of the system. The picture shows the overshoot at the low frequencies due to the resonance of the internal loop. The resonance frequency and amplitude is in facts depending on the loop gain, with a lower gain resulting in a lower frequency (kP=250, freq=5 Hz). From the plot we observe that the system shows in general an effective attenuation ($T_f/T_p <1$ ), with significantly poorer performances in the 60 Hz and 100 Hz bands. We investigated such regions with dedicated frequency sweep measurements. \\
        We repeated the test with  different loop gains and the result is shown in Fig.~\ref{fig.latt-atten3}. In the figure we plot the tilt values instead of the attenuation, so that they shall be compared with the black plot which is the tilt measured with the TS attached on the RB. The three plots are well superimposed, meaning that the attenuation, in this frequency band, does not depend from the loop parameters. The resonance at 100 Hz, in particular, is identical for the three plots, thus suggesting it cannot be originating from the internal loop.\\
        We then repeated the sweep measurement at different gaps, i.e. with the TS floating at different distances from the RB. The result is shown in Fig.~\ref{fig.latt-atten4}. Under the assumption the behaviour is due to the air coupling, we expected to observe a larger attenuation at larger gaps. The experimental data confirmed the hypothesis and we observed a progressive increase in attenuation changing the gap from 20 um to 200 um. We also observed that the resonance peak is higher al larger gaps. This point is consistent with the following hypothesis:  in the band 60 Hz to 90 Hz, a large air gap transmits less efficiently the  external excitation; at 100 Hz a resonance of the TS is excited and a larger gap is less efficient in damping it.\\
        Such scheme is just a qualitative description. A test in a vacuum chamber is required to fully understand the behaviour.
        
        \fig{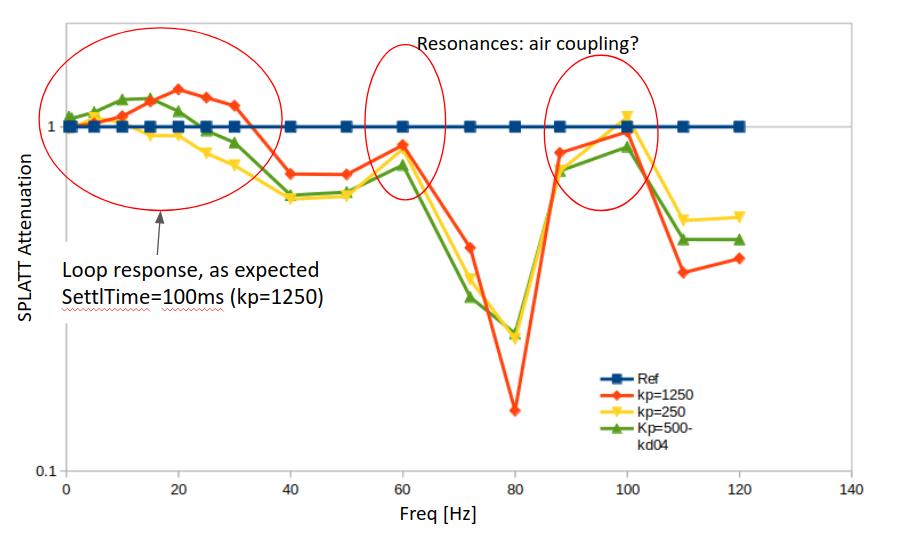}{9}{Relative frequency response as measured on the test bench. Blue line, reference, or the optical tilt measured with the TS pulled by the actuators toward the RB; red, green, yellow plots: $T_f/T_p$ as described in the text, measured with different actuator gains.}
       
        \figtable{latt-atten3}{Optical tilt measured with TS floating, different actuators gain, compared with the tilt measured with the TS pulled against the RB (black plot).}{latt-atten4}{Optical tilt measured with TS floating at different gap.}
        
    \section{The SPLATT simulation}\label{sec.simulation}
        \subsection{Simulation setup}
        The simulation toolkit is composed by three main parts: the DM, the PWFS and the closedloop.\\
        The DM is segmented and replicates the JWST geometry, with hexagonal segments on a hexagonal grid with 3 rings, within a circular outer mask. The segments alignment modes (piston and tip/tilt) are the system degrees of freedom (DoF) and are simulated by producing the associated shapes on the pupil mask. \\
        The PWFS and closedloop code are part of the PASSATA toolkit (see \citeonline{CITA5}), that has been used intensively for the simulation, design, and performance evaluation of the FLAO, ERIS, GMT, MAVIS AO system. In particular, PASSATA was adopted to simulate the WF sensing and control strategy (including the segments piston) for the GMT.        \\
        
        \noindent The first stage of the simulation, after importing the DM, is to calibrate the PWFS. This is done in two steps: as first the system mask is created by selecting those pixel with an illumination level above a user-defined threshold. As a second step, the PWFS interaction matrix is calibrated by measuring the PWFS signal when the DoF of the simulated  DM are excited individually. The measurement is differential according to the \emph{push-pull} technique: the commands are applied sequentially with positive, then negative amplitude, and the difference of the corresponding signals is taken. The command amplitude shall be chosen carefully to calibrate the system within its linear range.
        Our test model is composed by a segmented DM with 19 hexagonal segments and a circular mask, whose degrees of freedom are the local alignment modes, namely segment piston and tip/tilt; the PWFS images the DM on a 36x36 to 76 x76 pixel grid (per sub-pupil), in order to test the effect of a better resolution versus a lower photon signal per pixel. The WFS camera is the CCD39 (the one used for the FLAO\cite{flao} AO system at the Large Binocular Telescope), which has a known noise characteristic and is consistent with a worst case scenario (it is a quite old camera with important read-out noise and a quantum efficiency lower than 35 \%). A summary of the simulation parameters is reported in Tab.~\ref{tab.param}.
        
        \figtable{mask}{A sample of the DM pupil mask, segmented into 19 separated hexagonal sectors. The inter-segment gap is one of the parameters under test.}{ccdframe}{A typical PWFS raw frame: the pyramid tip is located at the geometrical center of the frame, while the 4 sub-frames are the pupil images produced by the pyramid 4 faces. The light unbalance amongst the faces is converted into a slope signal.}
        \fig{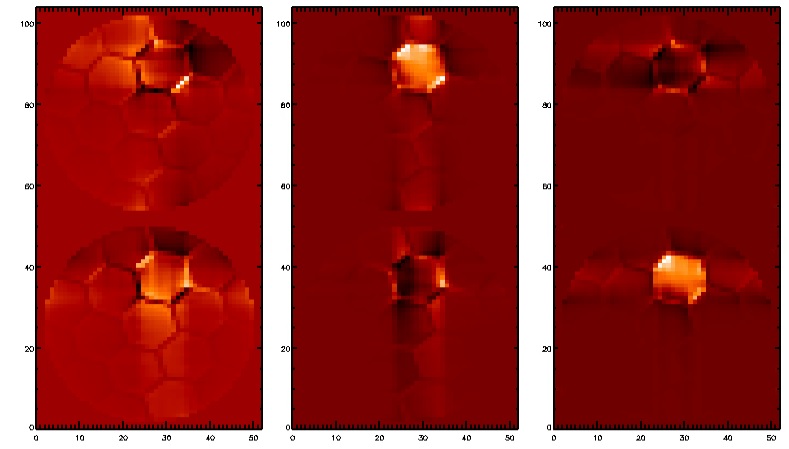}{9}{The slope signals corresponding to piston, tip, tilt (left to right) of segment \#1.}
        
        \subsection{Closed loop results}
        We tested closing the loop for a demonstration test case. We created an initial DM offset by scrambling the segments with piston and tip/tilt; such initial surface error is 50 nm RMS and is consistent with a preliminary co-alignment and co-phasing performed with a different device (the PWFS at low sensitivity or the scientific camera to identify and pre-adjust the single segments). We then closed the PWFS loop with the parameters reported in Tab.~\ref{tab.param}, in particular the guide star magnitude was 10 and the loop frequency was 10 Hz. The loop frequency is the PWFS frame rate; we must then consider that with a 0.1 gain the effective loop speed is 1 Hz. We are interested in the stability of the WF correction, which is related to the sensitivity of the PWFS. We therefore evaluated the loop performances by computing the dispersion of the DM surface error during the loop, after the initial convergence stage.  We basically computed the residual DM surface error versus time: we then calculated the standard deviation of the plot, as an estimation of the WF stability.\\
        The result (see Fig.~\ref{fig.closeloop1}) is very promising since we computed a sub-nanometer stability (or sensitivity) compared to a significantly fast loop speed and faint star magnitude.\\
        We are currently (summer 2022) running simulations to further asses these results, for instance by  evaluating the loop performances at different guide star magnitudes and  PWFS sampling, i.e. the number of pixel on each sub-pupil of the pyramid CCD.
        
        	\begin{table}[h]
	    \centering
	    \begin{tabular}{|c|c| c| c|} \hline
	    \multicolumn{4}{|c|}{\textbf{DM}} \\ \hline
	    DM diameter & 5.08 m to 5.19 m & N. Segments size & 19 \\ \hline
	    Segment geometry & hexagonal & Pupil geometry & Circular, un-obstructed \\ \hline
	    Segment orientation & 15 $\deg$ & Inter-segm. gap & 2 cm to 10 cm  \\ \hline
	    DM Pixel pitch & 5 mm  & Pixel on diameter & 140 \\ \hline \hline
	    \multicolumn{4}{|c|}{\textbf{PWFS}} \\ \hline
	    PWFS resolution & 36 to 76 pix & WFS $\lambda$ & 750 nm $\pm 150 nm$ \\ \hline
	    Frame rate & 10 Hz & Modulation amplitude & 0 $\lambda/D$ \\ \hline
	    CCD Camera & CCD39 & Noise & Photon, ReadOut \\ \hline
	    Quantum eff. & 0.32 & Binning & 1 \\ \hline
	    \multicolumn{4}{|c|}{\textbf{Control loop}} \\ \hline
	    Controlled modes & 56 & Type of control & Integrator \\ \hline
	    Delay & 1 & Gain & 0.1 \\ \hline
	     \multicolumn{4}{|c|}{\textbf{Parameters space}} \\ \hline
	    Guide star magnitude & 10 to 18 &  Modal amp. for calibration & 5 nm to 20 nm \\ \hline
	    Initial Surface Error & 50 nm RMS & Initial Surface error PtV & 250 nm \\ \hline
	     	    \end{tabular}	    \caption{Overview of the parameters to create the simulated active primary mirror.}
	    \label{tab.param}	
	   \end{table}

        \fig{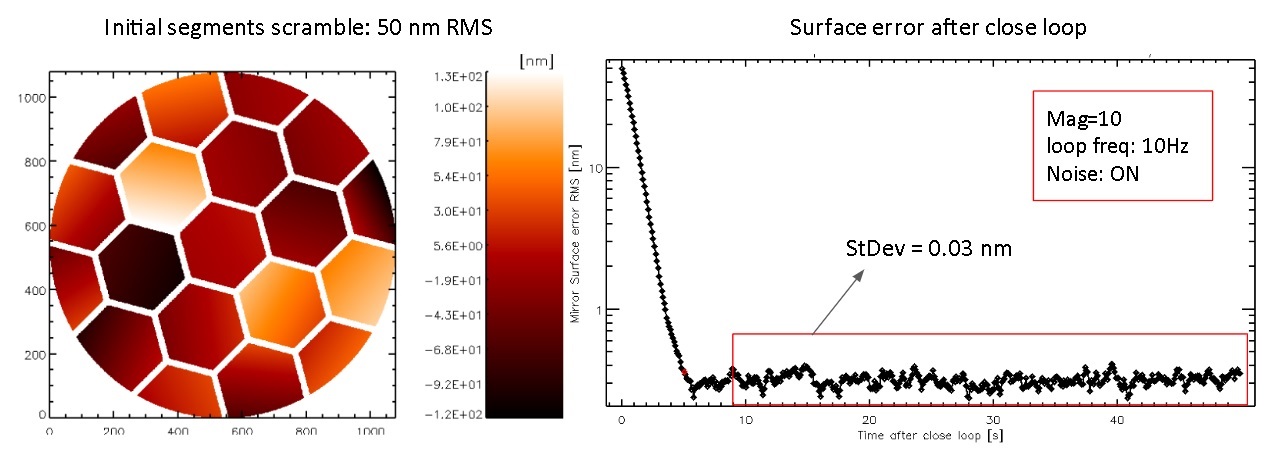}{6}{left panel: the initial surface offset, local piston,tip/tilt on the segments. Right panel: plot of the residual surface error on the DM, after closing the loop. The sample within the red box is taken as the stability of the PWFS.}
        
        \subsection{Discussion}
        Some points descending from the PWFS sensitivity need here to be discussed. As first, it would be possible to push the limit for the guide star magnitude toward the faint end. This implies a large sky coverage with the scientific target to be used also as a reference for the WFS. The case of the extended object shall be addressed to understand the possible limitations. The relatively fast frame rate implies that the open-loop stability of the DM (and of the entire system) can be limited to a sub-minute time scale. We could also consider another scenario (following a \emph{virtual adaptive optics} approach)  where the PWFS doesn't actually drive the DM, but its fast cadence readings are used for an enhanced data processing.

    \section{Conclusions and perspectives}
    The SPLATT project aims at investigating two key elements: contactless DM as segments of an active primary mirror, to exploit their intrinsic insulation from vibration coming from the payload; and sensitive PWFS to drive the correction chain at faster speed and even on faint reference stars. The first point has been addressed in the laboratory using the OBB, a 40 cm diameter demonstrator of active primary, with 19 actuators, manufactured and tested under an ESA TRP. The laboratory activity showed that the actuator local closed loop, controlling the position of the TS, behaves as a low pass filter and rejects effectively (at certain frequency bands) the external disturbances injected by shaking the test stand. The tests shall be repeated in vacuum to assess  the impact of the thin air gap between the TS and the RB, that could possibly induce vibrations on the optical surfaces.\\
    We then simulated a PWFS using a numerical code developed for ground based telescopes; we created a segmented primary mirror with 19 segments controlled in local piston and tip/tilt. The PWFS was verified at different resolution and closing the loop on guide stars of different magnitudes. The results are consistent with an expected very low noise propagation at the low spatial scale modes; we checked in particular the mirror WF scatter after convergence, taken as the measurement stability, and demonstrated a sub-nanometer sensitivity even with faint stars (magnitude 14) as references.\\
    The context of the project is the technological development for high stability optics and correction chain; contactless DM and PWFS could be a building block for the next generation of space telescopes.
    
    \section{Acknowledgements}
   The view expressed herein can in no way be taken
to reflect the official opinion of the European Space Agency. The LATT prototype is property of ESA and has been kindly made available by ESA for laboratory testing with a loan agreement.
The SPLATT project is funded by INAF - Istituto Nazionale di Astrofisica under the TECNO-PRIN INAF 2019 program.\\
In memory of Piero Angela, the most brilliant italian scientific journalist, who passed the day this paper was concluded. He inspired us all, when we were kids.

\newcommand{\procspie}{Proc. of SPIE}
\bibliographystyle{spiebib}

\end{document}